\documentclass[preprint,showpacs,aps,prc,groupedaddress,floatfix]{revtex4}
\usepackage{epsfig}
\usepackage{amsmath,amssymb}
\usepackage{bm}

\def\nuc#1#2{\relax\ifmmode{}^{#1}{\protect\text{#2}}\else${}^{#1}$#2\fi}
\begin{document}
\graphicspath{{figures/}}

\title{Elastic scattering phenomenology.}
\author{R. S. Mackintosh}
\email{raymond.mackintosh@open.ac.uk}
\affiliation{School of Physical Sciences, The Open University, Milton Keynes, MK7 6AA, UK}

\date{\today}

\begin{abstract}We argue that, in many situations,  fits to elastic scattering data that were historically, and frequently still are,   considered  `good',  are not  justifiably  so describable.  Information about the dynamics of nucleon-nucleus and nucleus-nucleus scattering  is lost  when elastic scattering phenomenology is insufficiently ambitious.  It is argued that in many situations, an alternative approach is appropriate  for  the phenomenology of nuclear elastic scattering of nucleons and other light nuclei. The approach  affords an appropriate means of evaluating folding models, one  that fully exploits available empirical data. It is particularly applicable for nucleons and other light ions.
\end{abstract}


\maketitle

\section{Introduction}
\label{intro}
In this note we argue that if elastic scattering phenomenology is worth doing, it is worth doing well;
moreover, it \emph{is} worth doing. When phenomenology is not done well, according to reasonable criteria,   opportunities to extract information from elastic scattering data will be lost.  However, rigorous phenomenology can uncover the information about nuclear interactions which is contained in high quality elastic scattering data, but which is often unexploited.  The appropriate fitting procedures are by no means new.  A key point is that the undulatory (wavy) potentials, that have sometimes been found  as a result of model independent fitting, are not aberrations,  but have a natural interpretation. For reasons that will be explained, the approach to phenomenology advocated here supports what is arguably the most appropriate means of evaluating folding models and other theories for optical model potentials.

The present discussion applies particularly to the elastic scattering of nucleons and other light nuclei and not in situations where elastic scattering is sensitive only to the potential in the far surface region.

\section{An example from alpha particle scattering}
\label{example}
Ref.~\cite{edf1} presents a thorough study of elastic alpha particle scattering from \nuc{40}{Ca}, \nuc{44}{Ca} and
\nuc{48}{Ca} using two very different models. One model is a fairly standard deep potential (Woods-Saxon squared)
and the other is a non-monotonic  potential based on an energy density functional (EDF) approach. A third model has a deep potential parameterised following  the non-monotonic form of the EDF potential. 

The exhaustive calculations yielded potentials that were, simultaneously, (i) profoundly different, and, (ii) fitted reasonably well according to the standards which are customarily styled `good fits'. 

It would indeed be interesting if it could be established that the EDF model is a sound basis for explaining nucleus-nucleus scattering, but since quite different potentials fit the data as well (or as badly) as EDF models,
nothing definite can be deduced from this work. We do not criticise the highly professional study of Ref.~\cite{edf1}
which is up to the generally expected standards for such work. In fact, the fits for wide angular range data
tended to have huge values of $\chi^2$ per degree of freedom ($\chi^2/{\rm DF}$), and for cases with a small 
angular range  $\chi^2/{\rm DF}$
was often not much lower. As usual in such studies, the large number of fits were presented in the 
published figures 
in such a way that `by-eye' judgment of quality of fit was impossible, the size of the experimental points typically
corresponding to a factor of two on the logarithmic differential cross section scale. That last  comment applies
to many publications in this field, and we shall not raise the point again.

In short, after much work, two contradictory potentials were presented, neither of which was justified by the quality of fit. 
We note here the similar work for alpha elastic scattering from Ni isotopes| Ref.~\cite{edf2}. For balance, note that the  analysis of  alpha-alpha elastic scattering in Ref.~\cite{edf3} appears to be exceptional.  In this case  the EDF model is applied in a case in which, at the lowest energies, the absence of absorption increases the sensitivity to the non-monotonic character of the model. The analysis does indeed appear to support the existence of a repulsive term in the strong overlap  radial region.

\section{Appropriate phenomenology}
\label{appropriate}
We quote from a very interesting article~\cite{AvOPA} the following: 
``Therefore, in order to avoid too much phenomenology in the description of these
data, numerous attempts have been made to replace the phenomenological real potential
of Woods-Saxon (WS) type by a more microscopic $\alpha$-nucleus potential using an effective
interaction. The double-folding (DF) model has become widely used with an effective
nucleon-nucleon (NN) interaction folded with the mass distributions of both the target
nucleus and the projectile.''   So, how do we apply phenomenology to simultaneously (i)  do justice to high quality elastic scattering data, and, (ii)  evaluate an elaborate folding model? Here we propose an answer.

Although Woods-Saxon potentials (and variants such as Woods-Saxon squared) still have a legitimate role to play in reaction calculations, they do not do justice to good quality elastic scattering data, and do not begin to extract all the physical information from such data.   Model independent fitting with sums of gaussians etc. leading to $\chi^2/{\rm DF} \sim 1$, more or less unprejudiced by theory, is much better. However, 
not knowing how to interpret  the results discourages such fitting; what do the resulting undulations mean? The contrast with analyses of electron scattering, which have become of great sophistication, see e.g. Ref.~\cite{SickT}, is conspicuous. This may be because the nuclear optical model potential, OMP, is seen as much more a \emph{model} concept than the nuclear or nucleon charge density. 

Theories of the OMP are well-developed, for recent folding models see Refs.~\cite{DTK15,egashira,kuprikov}, but  they are not easy to evaluate.   They are
generally based on a local density approximation, usually with no representation of shell effects, varying collectivity  or reaction channel coupling.  Current folding models yield potentials that are local and $l$-independent, with a smooth radial form. Yet there is a strong case that collective and other reaction processes lead to an $l$-dependent and dynamically non-local potential (exchange non-locality is reasonably well represented), see Ref.~\cite{ldep,KR90,RK90}. It is reasonable to expect  phenomenology to allow for the contribution of such processes but this is seldom the case when folding models are tested by fitting data. Generally, in such tests,  the potential is `corrected' either with  uniform renormalisation (bad), or by adding a model independent correction (better).   There may also be a Woods-Saxon~\cite{WS}  imaginary part with fitted parameters.

The uniform renormalisation of folding models is bad because it is intended to correct for inelastic processes, etc. But the contributions (dynamic polarization potentials, DPPs) from such processes are never smooth and the DPP is never uniformly proportional to the unrenormalized folding model potential. Therefore, renormalization will never provide a perfect fit to high quality elastic scattering data; it simply disguises the shortcomings of the folding model. It is, of course, just where the best models fail that we have a chance to increase understanding.
To clarify: there are certainly situations where Woods-Saxon potentials, and similar forms,  are appropriate; but there are also many situations where they are not.

\section{Suggested approach}
\label{suggest}
Phenomenology must recognize theoretical restrictions on what is possible; for example, the unitarity limit $|S_L|  \leq1$ must be respected. However, respecting this limit  does not exclude the existence of local radial regions where the potential is emissive; such regions  appear in DPPs arising from channel coupling.  Phenomenology should only respect   \emph{necessary} restrictions and the first step is to determine a suitable global potential if one does not exist. The second step involves scattering from the specific nucleus of interest.
\begin{enumerate} \item Employ a reasonable parameterized form to fit as wide a range of elastic scattering data
as is reasonable to define a global potential. Such potentials  exist, see  Ref.~\cite{KD} for nucleons, but usually do not have an unrestricted energy range.  At this stage, do not introduce `local' fits for specific ranges (e.g. around closed shells) as done, for good reasons, in Refs.~\cite{pang} for mass three.
\item Fit data for the specific nucleus with an additive correction to the global potential:  search on the parameters of a model-independent additive term, e.g.\ sum of Bessel functions, spline functions, gaussians, etc. The search should \emph{not} be restricted to smooth (non-wavy) potentials, and the lowest $\chi^2$ should be sought.  This, and the essential error analyses,  will be most meaningful where there exist data that are precise and which have a wide angular range.
\end{enumerate}
It is the resulting purely phenomenological potential for a specific target nucleus against which folding model theoretical potentials  for the same nucleus should be compared. This means of evaluating  the folding model  potentials should replace the direct fitting of experimental data by renormalizing or with additive potentials.  The procedure is not restricted to the evaluation of folding models, see for example  the first two of the following points:
\begin{enumerate}
\item The dependence of the OMP upon shell structure, upon the varying strength of reaction channels and upon the varying collectivity,  will be revealed naturally. For an early example of a link between collectivity and OMP parameters,  see Ref.~\cite{ghj}. 
\item The model independent fitting should \emph{not} avoid the possibility of some undularity (waviness) in the potential. Such undularity is known to arise from channel coupling (for references, see Section III of Ref.~\cite{ldep}), and therefore constitutes a source of information concerning reaction dynamics. Of course, some undularity might
be due to erroneous data, but the error analysis should help identify this.  For an example of where information was missed when model independent fitting was halted just where undulations would have appeared, see Ref.~\cite{ARF}.
\item Regarding the evaluation of folding models: fitting data by renormalizing the real potential while also fitting the parameters of a purely phenomenological imaginary potential is susceptible to interplay between the effects of changes in the real and imaginary potentials, particularly where the data are less than optimal. 
\item The suggested approach avoids the problem that can arise when evaluating certain folding models in which there is a consistency problem:  certain corrections depend on the potential itself, requiring iterations as the potential is renormalised. That problem is absent with the method suggested here.
\item The inadequacy and incompleteness of a data-set will be revealed by the error analysis of the first step. In fact, nucleon scattering data is virtually always incomplete owing to the absence of spin rotation (Wolfenstein's parameter)
measurements, see Refs.\cite{wolf,KMR,RM80}.
\item The phenomenological potential, found as above,  will be the local and $l$-independent  equivalent to a
non-local and  $l$-dependent potential (non-locality due to both exchange and dynamical processes). It can be compared with any theoretical local potential which should ultimately contain representations of all contributing processes. Comparison is possible since all $l$-dependent and non-local potentials have local equivalents with  the same $S_l$ or $S_{lj}$. 
\item Potentials with local emissive regions should not be avoided as long as $|S_L|  \leq1$; regions of emissivity can occur in $l$-independent equivalents of $l$-dependent potentials that closely fit data and also in potentials representing reaction or collective coupling. 
\end{enumerate}

None of the above de-values global potentials such as those of Refs~\cite{KD,pang}. Their
properties, such as energy dependence, provide information concerning reaction dynamics, and they are essential for reaction calculations when there is a lack of relevant elastic scattering data. However, global potentials do not represent all the information concerning nuclear interactions for specific targets and projectiles;  that information remains untapped in the best existing elastic scattering data.  Potentials derived from precise and comprehensive data, in the manner described above, not only enable  a rigorous test of folding models but also reveal aspects of nuclear reaction physics that would otherwise not be noticed.

Point 2 above referred  to undulations arising from channel coupling. There are many examples of undularity generated by coupling to reaction or collective channels; for recent light ion examples see Refs.~\cite{KR90,RK90,MKPRC85,km11,mk11} and  for heavier ions Ref.~\cite{R94}. The $l$-independent potential that  has the same $S$-matrix, $S_l$ or $S_{lj}$, as an $l$-dependent potential will have undulations and Ref.~\cite{R94}  includes an example for $^{16}$O scattering from $^{12}$C.  Although theory implies the $l$-dependence of OMPs, see Ref.~\cite{ldep},  there is currently no  `dictionary' relating specific forms of undulations, such as  might be found by model-independent data fitting,  to specific forms due to reaction coupling or due to phenomenological $l$-dependence.  Even without such a dictionary, it would clearly be of great interest if
radial forms similar to those found in Refs.~\cite{KR90,RK90,MKPRC85,km11,mk11} were found in precision fits.

We have discussed two methods of evaluating folding models (or any theoretical OMP): (i)  fit elastic scattering data with a model-independent function added to the folding model potential, and, (ii) compare the folding model potential with the optimum model-independent potential determined in the way we have proposed. While method (i) is clearly superior to fitting by applying a uniform normalisation, method (ii) has several advantages. Firstly, many variant folding model potentials can be evaluated against the same empirical potential and can also be mutually compared; there is no need for new searches as the folding model undergoes development. Furthermore, different folding model calculations will have a uniform comparison.  Secondly, the real part can be evaluated independently of the imaginary part so there is no possibility of changes to the imaginary part confusing comparisons of real folding model potentials. Finally, the prior step, the determination of the local phenomenological potential, will reveal trends related to shell effects and variations in channel coupling contributions, etc.

\section{General discussion}
\label{conc}
Why is it necessary to make a case for the extraction of all the information contained in hard-won experimental elastic scattering data? Historically,  when the liquid drop and compound nucleus models dominated nuclear physics,  it was surprising  that a potential model could even approximately fit data for nucleon scattering from complex nuclei. Very simple models, with parameters that varied in a regular way, gave good enough fits to show that they had some validity. It soon became possible to get `reasonable' fits with systematically varying parameters for a wide range of target nuclei and energies~\cite{PEH}. Subsequently, various folding models also gave what, in some contexts, would be considered `reasonable' fits to elastic scattering data. Now, many years later,  the belief that approximate fits to elastic scattering are sufficient has lingered on.  It was natural in an earlier era not to require $\chi^2/{\rm DF} \sim 1$ as a criterion for a satisfactory fit. As we have argued, in many contexts this is no longer the case. It is now possible to interpret, qualitatively at least,  the undulatory properties of potentials found with precise data fitting.

The arguments of this paper apply most strongly to the elastic scattering of nucleons and other light ions. However, they apply in certain cases of the scattering of heavier ions for which elastic scattering is sensitive over a considerable radial range. The sensitive range can be determined on a case by case basis, perhaps using a notch test.  For examples of where potentials for heavier ions have been established over a meaningful radial range  see Ref.~\cite{laRab}. That work  presents cases involving \nuc{16}{O} on \nuc{28}{Si} and \nuc{40}{Ca} as well as elastic \nuc{6}{Li} scattering on a range of nuclei. In  all of these cases wavy potentials emerge, indicative of strong channel coupling implying that corrections to relevant folding models will also be undulatory, ruling out uniform renormalization.

Precise data-fitting sometimes has a bad name, but without Kepler's,  Newton could not have
verified his theory from Tycho Brahe's data. Moreover Newton's theory was eventually replaced  by Einstein's  theory, supported by precise data-fitting revealing the precession of the perihelion of Mercury.  We conclude that there are scientific contexts in which precise data-fitting is worthwhile. In a situation where the latest folding models, such as Ref.~\cite{DTK15}, are $l$-independent it would surely be interesting if it could be shown by fitting data that the nucleon OMP is, in fact, $l$-dependent~\cite{ldep}. Although elastic scattering can give only indirect evidence for nuclear collectivity etc.~\cite{ghj},  elastic scattering data does typically have precision and angular range greater than that for inelastic and reaction channels.

Precise fitting of wide angular range elastic scattering data would enable, for the first time,  detailed evaluations of DPP calculations. Since the couplings responsible for the DPP actually lead to a non-local and $l$-dependent potential, the evaluation of the local equivalent would be a necessary step in determining how these processes affect reaction calculations based on the use of local OMPs.  The evaluation of DPP calculations should make it possible to relate elastic scattering to those properties of nuclei which do not vary smoothly with $N$ and $Z$.  It should also enable a more rigorous evaluation of that part of the folding model that is expected to vary more smoothly with $N$ and $Z$. Methods involving a uniform renormalisation do not do this reliably. As an example for heavier ions: \nuc{6}{Li} DPPs have been calculated  extending well inside the strong absorption region~\cite{pang1}; the measurement and precision fitting of elastic scattering for this case would enable a study of breakup dynamics.

\section*{Acknowledgements}
I am grateful to Nicholas Keeley for helpful comments on the manuscript and for many discussions over the years.

\end{document}